# Dye-Sensitized Nanostructured TiO$_2$ Film Based Photoconductor

Jani Kallioinen,[†,*] M. R. Hassan,[‡] G. S. Paraoanu,[‡] and Jouko Korppi-Tommola[†]

[†]Department of Chemistry
[§]Department of Physics
Nanoscience Center, University of Jyväskylä, P.O.Box 35, FIN-40014 University of Jyväskylä, Finland

## Abstract

Grooves were etched in a conductive layer of a conductive, transparent glass, and a nanoporous TiO$_2$ film was deposited on both the conductive and nonconductive area. The width of the grooves was 100 μm and 150 μm. A transparent TiO$_2$ film was dye-sensitized, covered with an electrolyte, and sandwiched with a cover glass. The conductivity of the dye-sensitized TiO$_2$ film permeated with electrolyte was studied in the dark and under illumination, and was observed to be dependent on light intensity, wavelength and applied voltage. This study shows that dye-sensitized nanoporous films can be used as a wavelength dependent photoconductor.

[*]Author to whom correspondence should be addressed. Email: jakallio@cc.jyu.fi, phone: +358-14 2602503 and fax: +358-14 2604756.

# 1 Introduction

A nanocrystalline, semiconductor film with large bandgap (> 3 eV) and small particle size is transparent in the visible wavelength region and has an enormous internal surface area to which a large number of dye molecules may covalently bind via linking groups.[1-3] In dye-sensitized devices, the attached dye molecules extend the photoresponse into the visible wavelength spectrum. Photon absorption on the surface of the nanoparticles by dye molecules causes electron injection from the excited state of the molecule into the conduction band of the semiconductor. At this point the electron and the hole are separated: the electron is in the conduction band and the hole in the oxidized dye. This process is ultrafast and initiates conversion of light into electricity in dye-sensitized solar cells (DSSC).[1-5] In DSSCs, following the charge separation, injected electrons are transported through the network of nanoparticles and the neutral dye is regenerated by a redox couple in the electrolyte. The redox species diffuse between electrodes and gain electrons at the counter electrode.

The photocurrent of dye-sensitized solar cells depends on the irradiance and illumination wavelength.[1-3] In photovoltaic applications, good light harvesting efficiency is required and therefore dye molecules should absorb in a wide wavelength spectrum and should have high quantum yield of electron injection and slow recombination kinetics.[3] To fulfill these requirements a wide range of dye molecules has been synthesized, characterized and tested in DSSCs. Under full sunlight an efficiency of 11 % has been reached by using highly purified Ru(2,2'-bipyridyl-4,4'-dicarboxylate)$_2$(NCS)$_2$(TBA)$_2$ based dye.[6,7] This type of ruthenium complex is used also in the present study. The dye has a wide absorption spectrum and photocurrent generation in photovoltaic applications starts already close to 800 nm.[2,6]

In the present study, the conductivity of a dye-sensitized film permeated with electrolyte was studied under applied voltage and illumination. The structure of our devices was closely related to DSSC, with the difference that we have no platinized counter electrode and that part of the conductive layer under the RuN3-TiO$_2$ film was removed.

# 2 Experimental

## 2.1 Preparation of samples

The schematic presentation of samples is shown in Fig. 1. Samples were made on 3 mm thick conducting, transparent glass (F-doped $SnO_2$, R = 15 Ω/square) with size of 10 mm times 15 mm. A line of conductive layer was removed either mechanically with a glasscutter knife (width of the gap ~150 μm) or with plasma etching (width of the gap was either 100 μm or 150 μm). Transparent anatase $TiO_2$ paste (Ti-Nanoxide T by Solaronix, average particle size 13 nm) was spread by doctor-blading. Part of the paste was on the nonconductive line and part on top of the conductive layer (see Fig. 1). After spreading, the film dried in air for a few minutes, and was heated up at ~450 °C for 30 minutes. The prepared film had thickness of 6 μm and size of 10 mm x 4 mm. The $TiO_2$ sample, still warm (~90 °C), was sensitized in 0.3 mM solution of ruthenium 535 bis-TBA [cis-bis(isothiocyanato)bis(2,2'-bipyridyl-4,4'-dicarboxylato)-ruthenium(II) bis-tetrabutylammonium; from Solaronix] in ethanol for 14 hours. After dye-sensitization, a plastic frame/spacer with a thickness of 130 μm was placed on top of the dye-$TiO_2$ film. To minimize direct reactions with the $SnO_2$:F layer, the open area of the frame was 8 mm x 2 mm, which was smaller than the area of the dye sensitized film. The empty volume in the frame was filled with an electrolyte: 0.1 M LiI (Aldrich, 99.9%), 0.05 M $I_2$ (Aldrich, 99.8%), 0.3 M tert-butylpyridine (Aldrich, 99 %), 0.5 M 1-hexyl-3-methylimidazolium iodide (synthesized according to reference 8) in 3-methoxypropionitrile (3-MPN, Fluka, 99%). The electrolyte is the same we have used previously in our studies for dye-sensitized solar cells.[9] A microscope glass was placed on top of the spacer frame and glasses were pressed together with clamps. Finally, silver paint was added on conductive layer to ensure good contact during measurements.

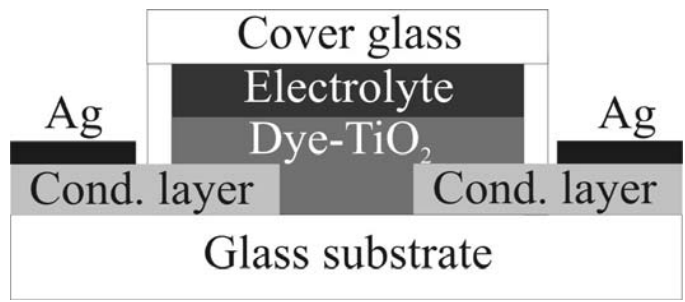

**Fig. 1.** Schematic presentation of the device.

*2.1.1 Patterning of the conductive layer by electron beam lithography and plasma etching*

The conductive layer on top of a glass substrate was patterned into two areas by using electron beam lithography and reactive ion etching. First, the conductive glass was covered with a resist sensitive to electrons bombardment, working as a protective layer against argon plasma etching. A thick layer of resist was required and thus a copolymer (Microchem MMA(8,5)MAA EL11 positive radiation sensitive resist, 11% in Ethyl Lactate) was used instead of standard PMMA polymer. The copolymer was spin-coated on top of the conductive glass (2000 rpm for 50 seconds). This was followed by heating on a hot plate at 160 °C for a few minutes to form a stable resist layer. The procedure was repeated three times to generate a resist layer with total thickness between 1.5 μm and 2 μm. A scanning electron microscope equipped with lithography software was used to draw a line into the resist layer. As the width of the samples was relatively large, ~10 mm, a low SEM magnification was used in line drawing. The parameters for the lithography were as follows: acceleration voltage 30 kV, dose 1, resist sensitivity 170 and beam current 100 pA. To remove the irradiated line we immersed the sample in a mixture of methanol and methylglycol (2:1) for 40 seconds. Etching of the conductive layer was done with reactive ion etching (argon plasma). The etching parameters were: argon flow 50 sccm and plasma power 200 W. The time required for complete etching of the conductive oxide layer was 40 minutes. Last, warm acetone was used to remove the resist and reveal the final structure. A measured infinite resistance between the two electrodes (across the groove) confirmed the etching result and SEM images showed uniformity of the groove. Some side effects such as slight burning (over heating) of the protective layer were observed.

## 2.2 Current-voltage measurements

A Keithley 2400 source measurement unit was employed in most of the studies as a voltage source. Measurements were controlled and data was collected with Labview 6.1 software. In voltage sweeps a bipotentiostat from Pine Instrument Company (model AFCBPI) was used.

## 2.3 Illumination

In studies with different light intensities, a tungsten lamp (Oriel Instruments) with a water filter was used together with an interference filter with center wavelength of 588 nm (FWHM = 12 nm). The irradiance was adjusted by changing the output power of the lamp. In spectral response measurements the same light source was used with 20 different interference filters in the wavelength region 404 nm - 808 nm, and FWHM varying from 10 nm to 14 nm. The intensity of light at each wavelength was measured with a calorimeter (Scientech). In current-voltage measurements a halogen lamp (Solux 12V/50W/4700K) was used to produce an illumination with 0.1 sun conditions. The irradiance was adjusted using a calibrated mono-Si-KG3 solar cell. In voltage sweeps white light with intensity of ~1000 lux was used (measured with INS illumination meter DX-200).

## 3 Results and Discussion

The prepared sample cell was illuminated at different wavelengths and the conductance was measured at a bias voltage of 0.5 V. Measured values were scaled with incoming light intensity; normalized data together with a Gaussian fit is shown in Fig. 2(b). The highest conductivity was attained close to 540 nm which is the maximum absorption of the dye on $TiO_2$, see Fig. 2(a). The measured wavelength response spectrum is very similar with the incident photon-to-current efficiency spectrum recorded for DSSC which was based on the same dye-$TiO_2$ system, see Fig. 2(a).[2,10] This shows that the wavelength dependency of photoconductivity is due to absorption of dye molecules. In dye-sensitized $TiO_2$ film, photon absorptions initiate electron injection from the excited dye molecules into conduction band states of $TiO_2$.[1-5] After the electron transfer, electrons are in the conduction band/trap states of $TiO_2$ and holes are localized in the dye cations. In nanostructured $TiO_2$ films, charge transport is not driven by electric fields[3,11] ; instead, the transport is diffusive, due to differences in electrochemical potentials (from high carrier concentrations to low carrier concentrations).[11] In our device, dye cations and ions in the electrolyte can also take part in the charge transport. The dye cations might recombine with conduction band electrons, gain electrons from the redox species ($I^-/I_3^-$) in the electrolyte or the hole can diffuse towards negative electrode. Furthermore it has been observed that concentration and type of ions influence charge transport in nanoporous $TiO_2$ films, at least in a film without dye-sensitization.[11] To resolve which of the above transport mechanisms are relevant in the studied device, further investigations on influence of electrolyte should be done.

We can conclude at this point that our samples' photoconductivity is dependent on the absorption of sensitizer dye. Thus the photoresponse region of the system could be tuned by changing the sensitizer molecules. This could allow tuning of the conductivity properties to certain color/wavelength region. In the present study other sensitizers were not tested, but several dyes have been widely studied in photovoltaic cells. Based on literature studies, photoresponse region could be tuned from blue to near-infrared simply by changing the dye molecules.[12-14]

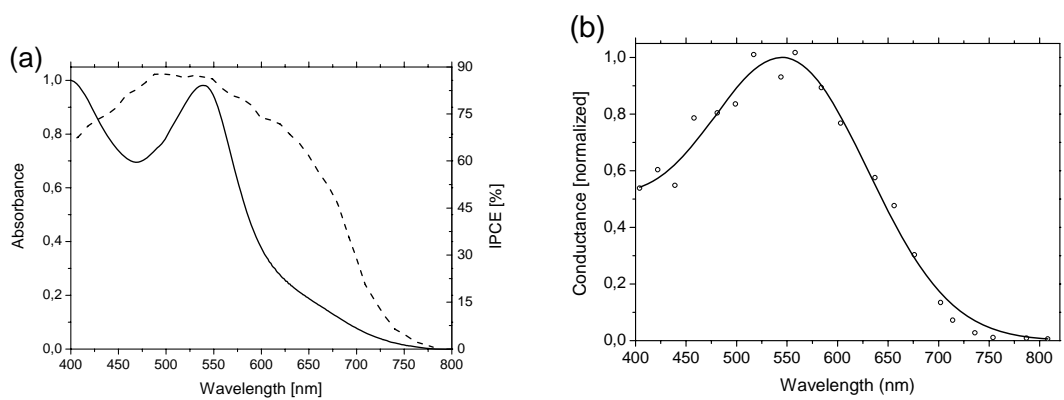

**Fig. 2.** (a) Absorption spectrum of dye-sensitized $TiO_2$ film in acetonitrile (solid line) and incident photon-to-current efficiency (IPCE) of DSSC (dashed line; data is taken from the reference 10) and (b): wavelength dependence of conductance of the studied device. Open symbols are data and solid line is a Gaussian fit. The highest conductance value is attained close to 540 nm, which is absorption band maximum of the dye.

It was observed that conductance of the sample is dependent not only on the illumination wavelength but also on photon flux into the photoactive area, i.e. the dye-sensitized film. This change of resistance/conductance at different irradiances was studied at a wavelength of 588 nm and the results are shown in Fig. 3. At low-light illumination (0.15 mW/cm$^2$) the resistance was ~1 MΩ and it decreased to 9 kΩ at an irradiance of 1.64 mW/cm$^2$. The conductance is quite linear in the measured irradiance region. When the illumination was done through the electrolyte, slightly lower conductance values were observed as a result of decreased light intensity through the electrolyte into the sensitized film (data not shown).

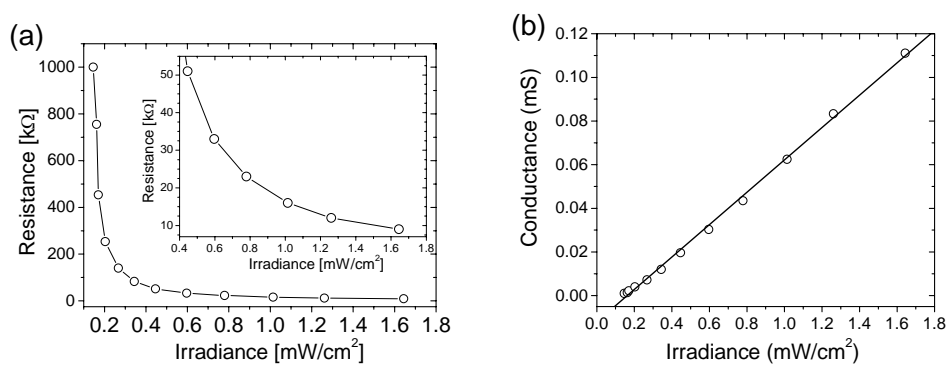

**Fig. 3.** (a) Resistance and (b) conductance as a function of irradiance at a wavelength of 588 nm (FWHM = 12 nm). The bias voltage was 0.5 V. Open circles are data and the solid line is a linear fit.

The above measurements were all made at bias voltage of 0.5 V. To study the effect of bias voltages on conductivity, voltages between 0 V and 1.2 V were swept under 1/10 Sun illumination and in the dark. In Fig. 4(a) we show the current-voltage curve of the device under illumination, and in Fig. 4(b) we show the corresponding dynamic resistance dV/dI as a function of voltage. The dark current of the device is close to zero until 0.6 V is reached. From 0.6 V the current slowly increases and starts to rise more clearly above 1.0

V. The low value for the current under voltage of 1.0 V indicates that the rate of reactions between the conductive layer and the electrolyte are very small. When the sample was illuminated, the current increases with increasing bias voltage, but not linearly. The current increase slows between 0.2 - 0.6 V, where the highest dynamic resistance (maximum at 0.36 V) is reached. Above 0.6 V the dynamic resistance decreases; the device reaches an equivalent static resistance V/I of 1.94 k$\Omega$ at a voltage of 1.20 V.

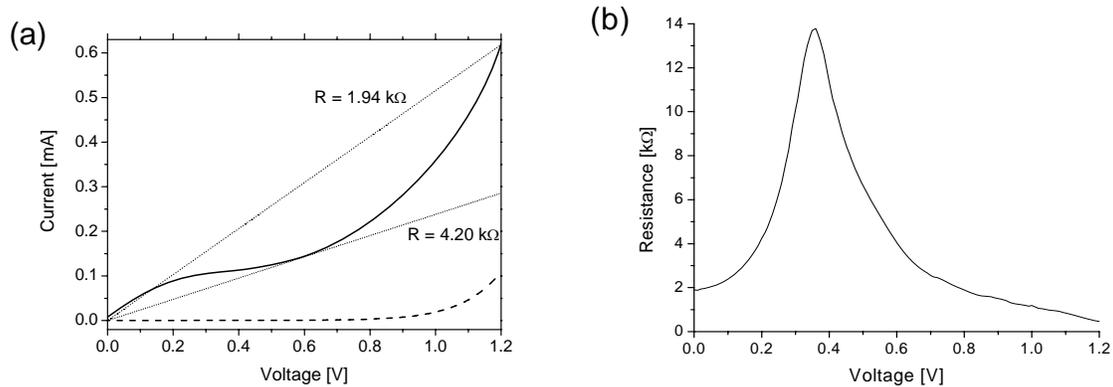

**Fig. 4.** (a) Current-voltage sweep of a photoconductor at illumination of 1/10 Sun (solid line) and in the dark (dashed line). Dotted lines represent equivalent static resistances of 1.94 k$\Omega$ and 4.20 k$\Omega$. (b) Dynamic resistance dV/dI at different voltages under illumination. The TiO$_2$ film had a thickness of 6 μm and the width of the nonconductive area was 150 μm.

Conductivity properties were also studied by sweeping several times between -1.2 V and +1.2 V, see Fig. 5. The sweep was started at -1.2 V and five cycles were measured in the dark and under low intensity white light. In the dark the current decreased from -1.2 V to -0.5 V and remained close to zero up to a voltage of +0.6 V, when it has started to rise. A hysteresis effect is observed in the voltage range 0.5 V to 1.1 V and in the same range at negative voltage. Under illumination, the current decreases rapidly from -1.2 V to -0.90 V, and then slowly (almost constant) down to -0.40 V. Zero current is reached at -0.05 V and there is a small positive current out of the cell at zero voltage. At positive voltages the current increases again slowly in the region of 0.40 V to 0.90 V. Under illumination the influence of the sweep direction is observed in voltages from -0.40 V to +0.40 V, otherwise the curves are overlapping; this hysteresis effect is also visible in the dynamic conductance at the same bias voltages, as shown in Fig. 5(b). At zero voltage the cell produces a small, non-vanishing current and the dynamic conductance reaches a local maximum. The sign of the current depends on sweep direction (scanning speed was of 200 mV/s); this indicates that under illumination and at low bias voltages the cell accumulates part of charges, which later get discharged at voltages higher than a threshold value (about 0.5V).

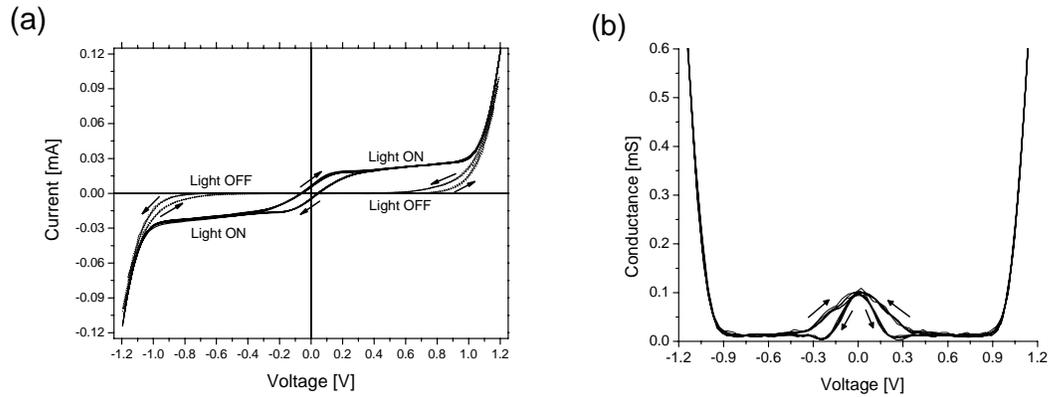

**Fig. 5.** (a) Current-voltage sweep in the dark (dotted line) and under low intensity illumination (solid line). Five cycles were measured between voltages of -1.2 V and +1.2V (speed of 200 mV/s). (b) Dynamic conductance dI/dV (overlap of several cycles) under illumination. The arrows show the direction of the scan.

Finally, the IV characteristics shown in Fig. 5 are remarkably similar to those of standard phototransistors, which hint to a potential application niche for these devices.

## 4 Conclusions

Electron conduction in the conduction band of dye-sensitized $TiO_2$ film is a known phenomenon in dye-sensitized solar cells under illumination. Here, we have studied how dye-sensitized nanoporous $TiO_2$ film can be used as a photoconductor. Under applied voltages (< 1 V), the resistance of the dye-$TiO_2$ film permeated with electrolyte decreases significantly when the illumination is switched on. At low-light illumination, the resistance of the device was of the order of MΩ, whereas at higher intensities it decreased down to a few kΩ, depending on the applied voltage, wavelength and irradiance. The conductance of the device was shown to be wavelength-dependent due to absorption properties of the dye molecules. This could allow tuning of the wavelength response region of the photoconductor simply by changing the dye.

## Acknowledgements

J.K. thanks Technology Agency of Finland (TEKES) for financial support. G. S. P. acknowledges EU SQUBIT-2 (IST-1999-10673), and support from the Academy of Finland (Acad. Res. Fellowship no. 00857, and projects no. 7111994, 7118122, 7205476), and Tekes/FinnNano MOME. M.H. was supported through CoE in Nuclear and Condensed Matter Physics JyU.